\documentstyle[aps,multicol,pre]{revtex}
 \title{Role of structural fluctuations in the insertion 
 into complex host matrices
}

 \author{E.V. Vakarin  and J.P. Badiali
}
 \address{LECA
 ENSCP-UPMC, 11 rue P. et M. Curie, 75231 Cedex 05, Paris, France\\
 }

\begin{document}
 \maketitle
 \begin{abstract}
 A coupling of structural and thermodynamic fluctuations in the course
 of various-type insertion processes is investigated within a combination
 of Gibbsian statistics and the information theory approach.
 It is shown that the coupling makes it possible to restore 
(at least partially) the information, inaccessible from experimental tests.
This enables one to make physically reasonable predictions under limited
information on the system.	
  \end{abstract}

\begin{multicols}{2}

\section{Introduction}
Intercalation processes find their application in many technologically 
important domains, such as the design of   
hydrogen-storage systems\cite{HS}, rechargeable high-energy  batteries,
electrochromic devices, (see Ref.\cite{rev} for a review), electroactive
polymers, and superconductors \cite{sup}. Microscopically the insertion is a 
complex process involving many effects (e.g. the charge transfer, elastic 
response of the matrix, the permselectivity, etc). Even well-characterized 
materials, like crystalline or layered compounds, exhibit rather complicated 
elastic properties, involving restructuring, staging and random distortion of
the galleries\cite{Mahanti}. For amorphous (or porous) matrices
\cite{Julien} the situation is complicated by the host heterogeneity, that
may change in the course of intercalation. For instance, structural 
heterogeneity of electroactive polymers changes because of the polymer
swelling. Complex geometry and poorly characterized energetics\cite{aerogel} 
make it practically impossible to construct a mechanical model. Therefore, 
statistical mechanical investigations in this domain face difficulties, 
induced by the system complexity. In this situation it seems reasonable to 
develop a new theoretical scheme, that would be capable of making 
reasonable predictions under restricted information. 
  
For this purpose we combine the standard statistical mechanics and the
information theory approach\cite{jaynes}. The latter has been introduced as 
an alternative (with respect to the standard statistical mechanics) tool for 
the statistical description of many-body systems.   
Such 
an approach has been successfully applied to a description of 
non-equilibrium steady states and the self-organization\cite{haken} in 
various complex (physical, chemical, biological, etc.) systems. One of the 
main advantages of the information theory is its capability of describing 
the equilibrium and nonequilibrium states in the framework  of a unified 
scheme, resembling the usual Gibbs statistics, provided that a suitable 
information (entropy) measure is chosen. On the other hand, the approach
does not rely upon microscopic details. 
This allows one to study the 
systems deviating from the standard equilibrium conditions. Quenched 
systems, like fluids adsorbed in porous media\cite{given}, spin 
glasses \cite{SK,Mezard} or liquid glassformers\cite{Pat} could serve as 
relevant examples.

In this paper we consider complex systems 
in which the information at 
a microscopic level is not sufficient for making a link between the system 
mechanics (Hamiltonian) and its statistics (probability distribution).
Then the we can split the system into subsystems with different levels of
description. One is a dynamic subsystem, which evolves according to a 
Hamiltonian, containing unknown parameters. The latter  reflect a 
coupling to a stochastic subsystem which is governed by a probability
distribution. This differs from the usual quenching in, at least, two 
aspects. The dynamic subsystem can influence the stochastic one and their
coupling determines both the thermodynamic behavior and the shape of the
probability distribution. For instance, by the analogy with the 
intercalation system\cite{PRB,JPCB,PRBd},  we can expect a change of the 
matrix volume upon the fluid absorption. Secondly, one necessarily deals 
with two entropic impacts: the thermodynamic entropy (due to the dynamic 
subsystems) and the information entropy (due to the probability 
distribution). Constructing from these two a suitable entropy measure,
we investigate the strain distribution and the guest thermodynamics in
homogeneous and heterogeneous matrices.   

\section{Maximum entropy estimation}
Let us assume that the states of a system are labelled by a continuous 
variable $x$, with $p(x)$ being the probability of a state $x$. 
Nevertheless, because of the system complexity, the probabilities are not 
known {\it a priori}. The only available information is on a  
constrained quantity 
\begin{equation}
\label{constr}
Q=\int dx p(x) q(x)
\end{equation}
which is the expectation value of a physical observable (e.g. energy, heat 
flux, etc.). In addition the probability distribution is assumed to be 
normalized
\begin{equation}
\label{norm}
\int dx p(x)=1
\end{equation}
Given this information can we estimate another relevant quantity 
$G=\int dx p(x) g(x)$? For this purpose one has to restore the probability 
distribution $p(x)$.

The basic role in this approach is played by the entropy $S$ that 
gives a measure of missing information\cite{jaynes} concerning the system 
state. If we have the entropy as a functional of the probability distribution
$S=S[p(x)]$, then, according to the information theory 
approach\cite{jaynes,haken}, the probability can be estimated through 
the entropy maximization under constraints (\ref{constr}),(\ref{norm}).
If we take the entropy (or the information) measure in the Shannon form
\begin{equation}
\label{shannon}
S=-\int dx p(x) \ln p(x)
\end{equation}
then such a variation procedure gives
\begin{equation}
p(x)=\frac{e^{-\gamma q(x)}}{Z};\qquad
Z=\int dx e^{-\gamma q(x)}
\end{equation}
where $\gamma$ is a Lagrange multiplier which has to be determined
from the constraint (\ref{constr}). For instance, if $Q$ is associated
with the equilibrium internal energy, then $p(x)$ becomes the conventional 
Gibbs distribution, with $\gamma=1/(kT)$.

\section{Complex systems} 
As is discussed above, by complexity we mean a situation when the 
available microscopic information on a system is not sufficient for its 
description in terms of a Hamiltonian. Thus, only one subsystem,     
say $\{s\}$, is governed by a Hamiltonian 
 $H[\sigma,s]$, while the rest (the surrounding) is 
 specified by a set of relevant parameters
$\{\sigma\}$ which appear with a probability distribution  $P(\sigma)$. 
Therefore, we have a coupling of a dynamic system to a stochastic one, which 
however influence each other. This means that $P(\sigma)$ is unknown,
the only available information concerns with an observable quantity
\begin{equation}
\label{expect}
E=\int d\sigma P(\sigma)E(\sigma) 
\end{equation}
that gives, for instance, a characteristic energy scale for the surrounding. 
In the case of spin glasses the
 counterparts of $\{s\}$ and  $\{\sigma\}$ systems are the spin variables
 and the random fields (or exchange constants), respectively. For porous 
 media,  $\{\sigma\}$ can be identified with the matrix (e.g. pore size)
 and $\{s\}$ - with the adsorbate degrees of freedom.

  For a given 
  configuration $\{\sigma\}$ one can calculate the partition function of the 
  dynamic subsystem 
  \begin{equation}
  Z(\sigma)=\int (ds) e^{-\beta H[\sigma,s]}
  \end{equation}
  where $\beta=1/(kT)$ is the inverse temperature. Then all 
  the thermodynamic characteristics are known. For instance, the 
  free energy 
  \begin{equation}
  F(\sigma)=-\frac{1}{\beta} \ln Z(\sigma) 
  \end{equation}
  and the internal energy 
  \begin{equation}
 \label{Usigma}
 U(\sigma)=-\frac{d}{d\beta}\ln Z(\sigma)
 \end{equation}
 determine the thermodynamic entropy
 \begin{equation} 
 S_T(\sigma)=\beta[U(\sigma)-F(\sigma)]
 \end{equation}
 Then the relevant thermodynamic quantities can be obtained by averaging
 over all realizations of $\sigma$ . For instance, the average free energy 
 is given by \begin{equation}
 F=\langle F(\sigma) \rangle=\int(d\sigma)P(\sigma)F(\sigma)  
 \end{equation}
 Note that, in contrast to the conventional equilibrium, now we have
 two entropies - the thermodynamic entropy
 \begin{equation}
 \label{STDN}
 S_T=\int(d\sigma)P(\sigma)S_T(\sigma)  
 \end{equation}
 and the one related to the information on the probability distribution
 \begin{equation}
 \label{SINF}
 S_I=-\int (d\sigma)P(\sigma)\ln P(\sigma)
 \end{equation}
For a given distribution $P(\sigma)$ the value of $S_T$ is the 
average thermodynamic entropy of the coupling $\{s\}$-$\{\sigma\}$. 
If, however, the
distribution is not known, then $S_T$ can be viewed as an additional entropy 
measure for inferring $P(\sigma)$ through the variation procedure.

In the case when the surrounding affects the dynamic subsystem, but not vice
versa (pure quenching) $P(\sigma)$ can be determined through the 
$S_I$ maximization under the constraint (\ref{expect}) and the 
normalization condition. This gives
\begin{equation}
\label{P0}
P(\sigma)=\frac{e^{-\gamma E(\sigma)}}{Z_0}
\qquad
Z_0=\int d\sigma e^{-\gamma E(\sigma)}
\end{equation}
where $E(\sigma)$ can be estimated through model or even scaling arguments.
For instance,  $E(\sigma)=a\sigma^2$ is the energy to create a field of 
the magnitude $\sigma$. As is discussed above, the Lagrange multiplier 
$\gamma$ should be determined from the condition (\ref{expect}). Once the 
probability distribution is determined, the thermodynamic quantities can be 
calculated straightforwardly.

Such a decoupling of the two subsystems is however an idealization, which
could be  more or less acceptable depending on a concrete situation.
In general, the surrounding should respond to the evolution of the dynamic 
subsystem. For instance, the porous materials could change the volume 
upon accommodation of the fluid, like in the case of intercalation
systems\cite{PRB,JPCB,PRBd}. Also, it has been demonstrated \cite{SUSC}
that the effect
of the quenching on the thermodynamics of $HCl$-ice interfaces weakens
with increasing ice film thickness.

 If the subsystems do influence each other, then the probability distribution 
must be evaluated starting from the total entropy $\Sigma(S_T,S_I)$, that 
describes the uncertainty concerning the state of the overall system. 
Obviously, $\Sigma(0,S_I)=S_I$ and $\Sigma(S_T,0)=S_T$ when there is no 
loss of information on one of the subsystems. 
The main problem is to construct $\Sigma(S_T,S_I)$ for nonvanishing $S_T$ 
and $S_I$. For instance, we can formally expand around the perfectly
ordered state $S_I=0,S_T=0$  
 
\begin{equation}
 \label{Stotal}
\Sigma(S_T,S_I)=\Sigma(0,0)+\kappa S_T+\lambda S_I+
 \delta S_T S_I+...
 \end{equation}  
 
As a first approximation the total entropy $\Sigma(S_T,S_I)$
can be estimated as a quasi-additive combination (we drop $\Sigma(0,0)$
as an irrelevant constant)
\begin{equation}
 \label{ST}
\Sigma(S_T,S_I)=S_I+\kappa S_T
 \end{equation}
where $\kappa$ is a parameter reflecting the coupling between the dynamic
and the stochastic subsystems.   
Then the distribution function can be inferred by maximizing (\ref{ST})
under the constraint (\ref{expect}). This leads to
\begin{equation}
\label{Pcoup}
P(\sigma)=\frac{e^{-\gamma' E(\sigma)+\kappa S_T(\sigma)}}{Z}
\qquad
Z=\int d\sigma e^{-\gamma' E(\sigma)+\kappa S_T(\sigma)}
\end{equation}
where the Lagrange multiplier $\gamma'$ should be determined from
the constraint (\ref{expect}). It is known \cite{fluct} that the entropy
$S_T(\sigma)$ determines the probability of thermodynamic fluctuations
for a given value of $\sigma$ according to
\begin{equation}
\label{fluct}
P_\varphi (\sigma)=\frac{e^{\kappa S_T(\sigma)}}
{\int d \sigma e^{\kappa S_T(\sigma)}}
\end{equation}
It should be emphasized that $P_\varphi (\sigma)$ does not describe
the fluctuation of $\sigma$, but the fluctuation of thermodynamic
variables related to $\{s\}$-subsystem (e.g. temperature, density, etc) for 
a given value of $\sigma$. Then the resulting distribution (\ref{Pcoup}) can 
be represented as \begin{equation}
\label{Pfluct}
P(\sigma)=P_\varphi (\sigma) \frac{e^{-\gamma' E(\sigma)}}
{\langle e^{-\gamma' E(\sigma)} \rangle_\varphi}
\end{equation}
where the average $\langle ...\rangle_\varphi$ is taken over the 
thermodynamic fluctuations of the dynamic subsystem. We thus see that
the statistics involves two ingredients: the "energetic", $E(\sigma)$
and the entropic $S_T(\sigma)$ or $P_\varphi(\sigma)$. On the other hand,
all thermodynamic quantities (like $S_T=\langle S_T(\sigma) \rangle$) are also
affected by these impacts.

\section{Strain distribution in the course of insertion}
In this section we analyze the interplay of these two impacts 
considering various aspects related to insertion systems.
In this case the guest plays the role of $s$-subsystem, whose
coupling to the host matrix is given by the matrix strain $\varepsilon$ 
which is a counterpart of $\sigma$-subsystem.
We are interested in determining the strain distribution $P(\varepsilon)$
and the guest thermodynamics $S_T$,
taking into account the host-guest coupling $S_T(\varepsilon)$. 
\subsection{Homogeneous matrix}
As an illustration of our approach we consider the simplest example
of guest insertion into a "homogeneous" host matrix. "Homogeneous" means
that the matrix properties (the structure, the binding site distribution, 
etc.) are similar for different matrix domains (e.g. graphite galleries). It 
is well-known that the matrix becomes strained with increasing guest 
concentration $x$. For a given $x$ the actual strain  is a sum of the domain 
strains $\sum_i \varepsilon_i$. For homogeneous matrices we may expect that
all the domains are equivalent, such that 
$\sum_i \varepsilon_i=n \varepsilon$, where $n$ is the number of domains.

Therefore, the host-guest coupling leads to a strain 
distribution $P(\varepsilon)$ and affects the guest thermodynamics. The 
strain-dependent guest entropy $S_T(\varepsilon)$ can be expanded around 
$\varepsilon=0$
\begin{equation}
S_T(\varepsilon)=S_T(0) +\alpha n \varepsilon +...
\end{equation}
Here $S_T(0)$ is the guest entropy in the case when the matrix is not
strained. This corresponds to the lattice gas model\cite{rev} 
describing the guest adsorption on a rigid lattice.
Note that all the quantities above ($S_T(0), \alpha, \varepsilon $) 
are concentration dependent.
The strain distribution $P(\varepsilon)$ can be determined by  
maximizing $S_I+\kappa S_T$ under the constraint that the
average strain 
\begin{equation}
\label{strain}
E=n \langle \varepsilon \rangle = n \int d \varepsilon P(\varepsilon) 
\varepsilon \end{equation}  
is known (e.g. measured experimentally). It can be 
estimated as the sample volume dilatation (or $c$-axis expansion for layered 
compounds).

According to eq.~(\ref{Pfluct}) 
we arrive at
\begin{equation}
P(\varepsilon)=\frac{e^{-(\gamma'-\kappa \alpha)\varepsilon}}
{\int d \varepsilon e^{-(\gamma'-\kappa \alpha)\varepsilon}}
\end{equation}
Determining $\gamma'$ from the constraint (\ref{strain}) we obtain 
\begin{equation}
P(\varepsilon)=\frac{e^{-\varepsilon/E}}
{\int d \varepsilon e^{-\varepsilon/E}}
\end{equation}
It should be noted that exactly the same distribution would be recovered
if only $S_I$ is used as the entropy measure. This is a consequence
of what we called homogeneity, that is the average $E$ is a good
estimation of the actual strain $\varepsilon$, which almost does not 
fluctuate. Then the experimentally available information (\ref{strain})
is relevant to the thermodynamic functions of interest. In other words,
if the experimental data allow one to recover an "exhaustive" picture of 
the system, then any refinement of the entropy measure (like introducing
$S_I+\kappa S_T$ instead of $S_I$) gives no new information.

Based on the probability distribution the guest thermodynamics can be 
predicted using the experimental data on the concentration dependence of 
the strain $E=E(x)$
\begin{equation}
S_T=S_T(0)+\alpha E
\end{equation}
Having determined $S_T$, we can recover all the
thermodynamic functions in the standard way. The free energy is also
linear in the strain (for simplicity we take the same coefficient $\alpha$):
$F=F(0)+\alpha E$. Based on this we obtain the guest chemical potential
as
\begin{equation}
\label{muh}
\mu=\mu(0)+\alpha \frac{dE}{dx}+E \frac{d \alpha}{dx}
\end{equation}
 This results coincides with the one obtained in our earlier studies
 \cite{JPCB,PRBd} combining the lattice gas model with the continuum
 elasticity theory. The latter implies that the stress can be defined as 
 $\alpha=dF/dE$.  
 If the stress $\alpha$ is composition independent,
 then from (\ref{muh}) we recover the well-known result of Larche and Cahn 
 \cite{Larche}.
\subsection{Heterogeneous matrix}
In reality the host matrices are not homogeneous. Even for materials 
which are conventionally regarded as well-characterized,  the heterogeneity 
is practically unavoidable in the experimental conditions (e.g. because
of defects, domain structures, etc). On the other hand the amorphous or
porous matrices are essentially heterogeneous and their properties are
known only statistically. In this case it is difficult to "invent" a simple
experimental test (like (\ref{strain})), giving an exhaustive description.
We assume that the insertion of guest species into a heterogeneous 
matrix results in a strain distribution for various matrix domains.
For simplicity we do not consider spatial strain correlations. This means
that the domain strains $\varepsilon_i$  are randomly 
distributed according to $P(\{\varepsilon_i\})=\prod_i P(\varepsilon_i)$ 
independently of the domain position $i$.

From the experimental point of view (e.g. measuring the matrix strain $E$ or
from another suitable test) we know only the range in which the individual
$\varepsilon_i$ can vary. The range of 
$\varepsilon_i$ variation should correlate with the average $E$. For 
instance, if $E=0$, then $\varepsilon_i$ are symmetrically distributed 
around $0$. For concreteness we choose: $0<\varepsilon_i<g_i(E)$.
Then, maximizing
$S_I$ under the normalization condition for $P(\varepsilon_i)$, we obtain
the following step-wise distribution
\begin{equation}
\label{Pexp}
P_0(\varepsilon_i)=\frac{H(\varepsilon_i)H(g_i-\varepsilon_i)}
{g_i}
\end{equation}
where $H(x)$ is the Heaviside step function. The distribution
$P_0(\varepsilon_i)$ is the one we can recover using the experimentally
available information. 
 
The guest thermodynamics can be predicted as before
\begin{equation}
\label{Stexp}
S_T=S_T(0)+\sum_i \alpha_i \langle \varepsilon_i \rangle_{0}
\end{equation}
\begin{equation}
\label{muexp}
\mu=\mu(0)+\sum_i\alpha_i \frac{d}{dx}\langle \varepsilon_i \rangle_{0}
\end{equation}
where $\langle \varepsilon_i \rangle_{0}=g_i/2$ and $\alpha_i$
is assumed to be composition independent. Note that the results above
are based on the estimation of the internal strains 
$\langle \varepsilon_i \rangle_{0}$, using the experimentally available
information on the "external" (observable from outside) strain $E$.  
As a simple approximation we may suppose $g_i(E)=f(S_i)E/n$, that is,
each $\langle \varepsilon_i \rangle_{0}$ contributes to the observed
strain $E$ proportionally to the domain size $S_i$, with
$\sum_i \langle \varepsilon_i \rangle_{0}=E$. 

These results can be refined
if we explicitly take into the host-guest coupling and determine
$P(\varepsilon_i)$ using $S_I+\kappa S_T$. This leads to the distribution
\begin{equation}
\label{Pnew}
P(\varepsilon_i)=\frac{e^{\kappa \alpha_i \varepsilon_i}}
{\int\limits_0^{g_i} d \varepsilon_i e^{\kappa \alpha_i 
\varepsilon_i}} 
\end{equation}
which is more informative in the sense that now not all values 
of $\varepsilon_i$ are equally probable. The average
\begin{equation}
\langle \varepsilon_i \rangle=
\frac{1+e^{\kappa \alpha_i  g_i}(\kappa \alpha_i  g_i-1)
}
{\kappa \alpha_i  (e^{\kappa \alpha_i  g_i}-1)}
\end{equation}
is different from $\langle \varepsilon_i \rangle_{0}$ obtained above. 
It is 
instructive to analyze the origin of this difference. For this purpose we
expand $\langle \varepsilon_i \rangle$ in terms of the coupling $\kappa$
\begin{equation}
\langle \varepsilon_i \rangle=\langle \varepsilon_i \rangle_{0}
+ \alpha_i \kappa \Delta_i^0
+... 
\end{equation}
where 
$$\Delta_i^0=\left[
\langle \varepsilon_i^2 \rangle_{0}-\langle \varepsilon_i \rangle_{0}^2
\right]=g_i^2/12$$ 
describes the strain fluctuations. The latter
are related to the domain size fluctuations and thus can be called the 
{\it structural fluctuations}. In the case of homogeneous matrices all
the domains are equivalent, then the structural fluctuations vanish 
$\Delta_i^0=0$ and 
$\langle \varepsilon_i \rangle=\langle \varepsilon_i \rangle_0$.   
Therefore, for heterogeneous matrices our estimation of the strain
statistics becomes more sensitive to structural fluctuations if we change
the entropy measure. This reflects the fact that for heterogeneous matrices
the actual strain $\overline{E}$ is different from the one ($E$) measured
as the sample dilatation. Their difference 
\begin{equation}
\overline{E}-E=
\sum_i[\langle \varepsilon_i \rangle-\langle \varepsilon_i \rangle_{0}]
=\sum_i \alpha_i \kappa \Delta_i^0
\end{equation}
takes into account the internal deformations which are not detectable
experimentally. This result is quite general, it does not
rely upon the simple form for $P_0(\varepsilon_i)$ and $P(\varepsilon_i)$ 
discussed here as examples. The only essential fact is that these two are 
different (compare (\ref{P0}) and (\ref{Pfluct})). The difference has its 
physical origin in the {\it mutual coupling} of the guest thermodynamics and 
the host structure. The coupling makes it possible to "restore" 
(at least partially) the information inaccessible from experimental tests. 
 
Based on the refined distribution (\ref{Pnew}) we can 
calculate the guest thermodynamics. 
\begin{equation}
\label{Stnew}
S_T=S_T(0)+\sum_i\alpha_i \langle \varepsilon_i \rangle_{0}+
\kappa \sum_i\alpha_i^2 \Delta_i^0+...
\end{equation}
Considering $S_T$ as a measure of the guest thermodynamic fluctuations
(see above), we conclude that the latter are coupled to the host structural
fluctuations. This reflects the fact that the matrix response to the guest
insertion is not only due to the host nature but also involves the host- 
guest coupling. Comparing the chemical potential 
\begin{equation}
\label{munew}
\mu=\mu(0)+\sum_i\alpha_i \frac{d}{dx}\langle \varepsilon_i \rangle_{0}
+\kappa \sum_i\alpha_i^2 \frac{d}{dx} \Delta_i^0
\end{equation}
with  (\ref{muexp}), obtained using the experimental information,
we see that (\ref{munew}) gives a more precise estimation, but also 
contains essential physics. In particular,
if $\sum_i \langle \varepsilon_i \rangle_{0}\to 0$, then from (\ref{muexp}) 
one would conclude that the insertion is almost topotactic and that the 
internal host distortion is irrelevant. In contrast, eq.~(\ref{munew}) 
implies that although the matrix is not strained on average, the expanded 
and contracted domains do contribute to the guest energetics. This is 
observed experimentally\cite{Julien} on various disordered host matrices, 
which exhibit negligible volume expansion in comparison to well-structured 
materials. Nevertheless, the electrochemical response of disordered
insertion systems is remarkably different. This can be understood in terms
of two effects. First is the configurational transitions\cite{JPCB} 
described by $\mu(0)$. Structural disorder induces a broad distribution
of binding energies on different matrix sites. This leads to quite steep
insertion isotherms with multiple short plateaus. A simple model capable
of recovering this effect can be found in ref.~\cite{icehet}. 

The second effect is due to the structural 
transitions, which are supplemented by structural fluctuations 
(see eq.~(\ref{munew})).  
In order to analyze this contribution within  our simple model, 
let us recall that $\langle \varepsilon_i \rangle_{0}=g_i(E(x))/2$ and
$\Delta_i^0=g_i^2(E(x))/12$. Assuming the simplest non-trivial 
concentration dependence $g_i(E(x))=b_ix$, we obtain from (\ref{munew}):
\begin{equation}
\mu=\mu(0)+\delta \mu +Kx
\end{equation} 
where $\delta \mu=\sum_i b_i \alpha_i/2$ induces a shift of the isotherm
in comparison to the strain free state, and 
$K=\kappa \sum_i \alpha_i^2 b_i^2/6$ plays the role of additional
repulsive ($K>0$) interaction. Since the chemical diffusion coefficient
is proportional to $d \mu/ dx$, it is clear that the structural
fluctuations tend to increase diffusivity in comparison to the strain free 
state.  This seems to be a general tendency for disordered matrices, as it 
follows from experimental studies\cite{Julien,porous}.

 \section{Conclusion}
 The coupling of structural and thermodynamic fluctuations in the course
 of various-type insertion processes is investigated within a combination
 of Gibbsian statistics and the information theory approach. The main focus
 is on complex (heterogeneous) insertion systems, where 
 the microscopic information is not sufficient for making a link between 
 the system mechanics (Hamiltonian) and its statistics (probability 
 distribution). Then the system can be viewed as an entropic coupling of
 a dynamic and a stochastic subsystem. It is shown that the coupling makes 
 it possible to restore (at least partially) the information, inaccessible 
 from experimental tests. This enables one to make physically reasonable 
 predictions under limited information on the system.	

It should be emphasized that our results are also applicable to matrices
which are initially homogeneous but become heterogeneous in the course
of insertion. For instance, this takes place in $Li$-graphite intercalation
compounds, which exhibit staging phenomena. In this case the occupied 
and empty graphite galleries are not equivalent. Also, if a gallery is 
partially occupied, then one deals with distorted domains\cite{Mahanti}.  
In this case the strain fluctuations as well as their spatial 
correlation should be taken into account. In a simple approximation this 
physics is reflected by our result (see eq.~(\ref{munew})). Our approach
is also capable of describing the swelling of electroactive polymers
in the course of intercalation. We plan to address these issues in a
future study. 

 
 
\end{multicols}
\end{document}